\documentclass[pss]{wiley2sp} 
\usepackage{amsmath}

\newcommand{\br}{\vec{ r}}

\begin{document}

\title{Atomistic modeling of dynamical quantum transport}

\titlerunning{Atomistic modeling of dynamical quantum transport}

\author{%
  Christian Oppenl\"ander\textsuperscript{1}, Bj\"orn Korff\textsuperscript{2}, Thomas Frauenheim\textsuperscript{2} and Thomas A. Niehaus\textsuperscript{\Ast,\textsf{\bfseries 1}}
}

\authorrunning{Thomas A. Niehaus}

\mail{e-mail
  \textsf{thomas.niehaus@physik.uni-regensburg.de}}

\institute{%
  \textsuperscript{1}\,Department of Theoretical Physics, University of Regensburg, 93040 Regensburg, Germany \\
   \textsuperscript{2}\, Bremen Center for Computational Materials Science, Am
  Fallturm 1a, 28359 Bremen, Germany\\
}

\received{XXXX, revised XXXX, accepted XXXX} 
\published{XXXX} 


\abstract{
We present dynamical transport calculations based on a tight-binding
approximation to adiabatic time-dependent density functional theory
(TD-DFTB). The reduced device density matrix is propagated through the
Liouville-von Neumann equation.  For the model system, 1,4-benzenediol
coupled to aluminum leads, we are able to confirm the equality of the
steady state current resulting from a time-dependent calculation to a
static calculation in the conventional Landauer framework. We also
investigate the response of the junction subjected to alternating bias
voltages with frequencies up to the optical regime. Here we can
clearly identify capacitive behaviour of the molecular device and a
significant resonant enhancement of the conductance. The results are
interpreted using an analytical single level model comparing the
device transmission and admittance. In order to aid future
calculations under alternating bias, we shortly review the use of
Fourier transform techniques to obtain the full frequency response of
the device from a single current trace.

%
%
%
%


{
}
}

%
%

\maketitle   
\section{Introduction}
The field of quantum transport at the molecular scale significantly
diversified over the last years \cite{Nitzan2003,Heath2009,McCreery2013}. While the interest was initially to
measure the conductance across individual molecules in an accurate and
reproducible fashion, current topics involve spin transport \cite{Bogani2008}, molecular
transistors \cite{Song2009}, thermoelectric effects \cite{Reddy2007,Nikolic2012}
and device heating \cite{Gagliardi2008,Schulze2008}. On the
theoretical side much progress was achieved using Green's function
methods in the energy domain \cite{Cuevas2010}. Time domain methods, on the contrary,
promise easy access to dynamical properties, like {\em ac} transport,
light-induced effects and higher harmonics in the current \cite{Chen2012}. In this
contribution we report on results of such a method based on
approximate time-dependent density functional theory, termed
TD-DFTB \cite{Niehaus2009,Wang2011}. The scheme allows to perform dynamical transport simulations
of realistic devices taking the electronic structure of molecule and
leads into full account. Extending an earlier study on a similar
topic \cite{Yam2011}, we first ask the question whether time and energy domain
methods provide the same answer for the steady state {\em dc} current. We
continue with a discussion of alternate currents and focus here
especially on resonant enhancement of the admittance beyond the low
frequency regime commonly studied.
\section{Method}
\begin{figure}
  \includegraphics[scale=0.38]{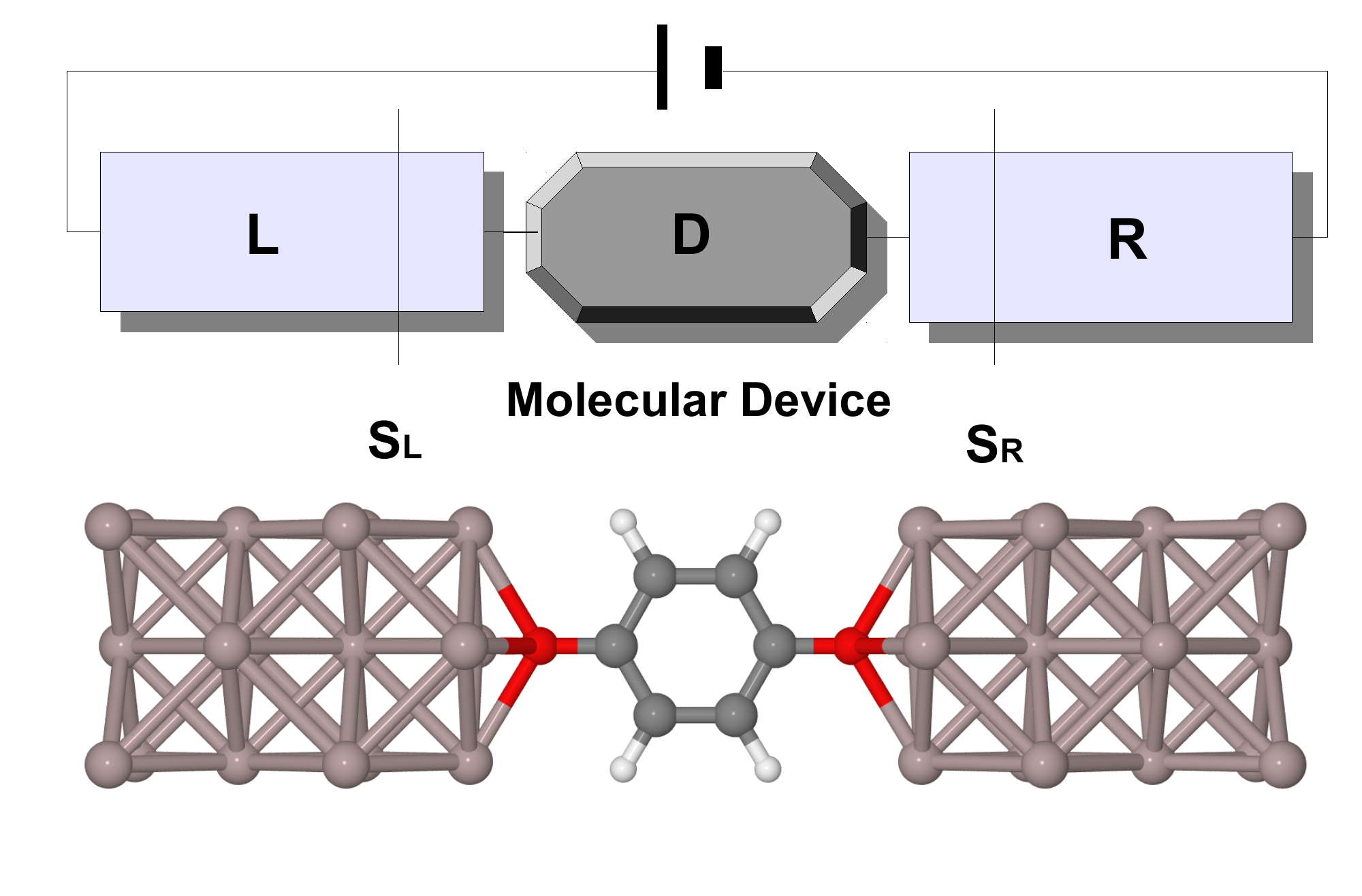}\\
  \caption{Schematic setup of the molecular device shown together with
    our test system. Only the atoms in the device region are shown.}\label{setup}
\end{figure}

In the following we present a brief description of our simulation method. A more detailed derivation and justification of the present scheme may be found in the original articles \cite{Zheng2007} and \cite{Wang2011}.
We assume a setup of the  molecular electronic device as depicted in
Fig.~\ref{setup}. The periodic left (L) and right (R) lead extend to
infinity and are in thermal equilibrium at the chemical potential
$\mu_{\alpha=L,R}$ with $\mu_L = \mu_R$ at $t=0$. At $t>0$, a
time-dependent bias potential $V(t)$ is applied that drives the
central device region (D) out of equilibrium and leads to a
time-dependent current. Instead of working with the full infinite
system, one can derive a Liouville equation for the device region only
\cite{Zheng2007} (in atomic units):
    \begin{equation}
    \label{neumann}
      i\frac{\partial}{\partial t}\sigma(t) = [H(t),\sigma(t)] -i \sum_{\alpha=L,R} Q_\alpha(t).
    \end{equation}

Here ${\sigma(t)}$ denotes the one-particle density matrix for the device region in a basis of localized atom-centered basis functions $\phi_\mu(\br)$. The Hamiltonian $H(t)$ is given in the adiabatic approximation of TDDFT \cite{Casida1995} and depends on the electron density $\rho(\br,t)$, while $Q_\alpha(t)$ incorporates all effects due to the metallic leads, especially also dephasing and dissipation. Numerically tractable and explicit forms for this term can be obtained from non-equilibriums Green's function theory in the wide band limit (WBL). As shown by Zheng et al. \cite{Zheng2007}, $Q_\alpha$ then takes the form:
\begin{equation}
Q_{\alpha}(t)=i[\Lambda_{\alpha},\sigma(t)]+\{\Gamma_{\alpha},\sigma(t)\}+K_{\alpha}(t),\label{WBLQ}
\end{equation}
where $\Lambda_{\alpha}$ describes the change of the device energy
levels due to the presence of lead $\alpha$, while $\Gamma_{\alpha}$
renders the lifetime of the molecular levels finite.\footnote{With
  respect to the
  article by Zheng et al. \cite{Zheng2007}, the designation of $\boldsymbol{\Gamma_\alpha}$ and
  $\boldsymbol{\Lambda_\alpha}$ is interchanged here. Square and
  curly brackets indicate a commutator and anti-commutator,
  respectively.} Both matrices are evaluated from first principles and
depend on the device-lead interaction and the lead surface density of
states. The term $K_\alpha$ involves only known quantities besides the
time-dependent bias potential $V(t)$ and hence Eq.~\ref{neumann}
represents a closed equation that can be numerically integrated by
conventional Runge-Kutta methods. To this end, the initial density
matrix at $t=0$ may be obtained without further approximations from
equilibrium Green's function theory in the WBL. As also shown in
reference \cite{Zheng2007}, knowledge of $Q_\alpha(t)$ allows one to
compute the time-dependent particle current $I_\alpha(t)$ through the left or right device-lead interface according to 
  \begin{equation}
I_{\alpha}(t)=-\text{Tr}[Q_{\alpha}(t)].\label{current}
\end{equation}  

In practical simulations the time step has to be chosen in the
attosecond regime in order to resolve the electron dynamics
accurately. This limits the accessible device dimensions and total
simulation time significantly. We therefore adapted the scheme from
above to the time-dependent density functional based tight-binding (TD-DFTB) method \cite{Frauenheim2002,Niehaus2009,Wang2011}. In essence, the TDDFT Hamiltonian matrix $H_{\mu\nu}(t)$ is replaced by 
\begin{eqnarray}
H_{\mu\nu}(t) &=& \langle \phi_\mu | H[\rho_0]|\phi_\nu \rangle
\\\nonumber &+& \frac{1}{2}[\delta V_A(t)+\delta
V_B(t)]S_{\mu\nu},\quad \mu\in A, \nu \in B.\label{DeltaV}
\end{eqnarray}  

The first term on the right hand side is the DFT Hamiltonian evaluated
at a time independent reference density $\rho_0=\sum_A \rho_A$, taken
to be a sum of atomic densities $\rho_A$ for each atom in the device
region. These densities and hence also the matrix elements can be
computed beforehand. The second term involves the overlap $S_{\mu\nu}$
of the basis functions and takes the deviation of the electrostatic
potential from the reference into account. The potential $V_A$ on the
atoms in the device region is computed at each time step from a
Poisson equation with boundary conditions determined by the given bias
potential in
the leads. The charge density required in this process is computed from the
density matrix $\sigma(t)$ \cite{Wang2011}. Besides this adaption in
the Hamiltonian, we follow the 
formalism of Zheng et al. without further modifications.        

\section{Results}
\subsection{Approach to steady state}
We applied the TD-DFTB scheme to the junction depicted in
Fig.~\ref{setup}. The 1,4-benzenediol molecule was optimized with
passivating hydrogens in
vacuum at the DFTB level and then symmetrically positioned inbetween Al nanowires of finite cross
sections. The device region consists of
the molecule and 36 additional Al atoms, while the simulation cell for the
leads included 72 atoms. The latter is periodically replicated to
$+\infty$ and $-\infty$ for the right and left lead, respectively, in
order to compute the surface Green's function and WBL parameters (see
Eq.~\ref{WBLQ}) at zero bias. The basis set is given by one s-type
atomic orbital for H and one s-type and three p-type orbitals for the other elements. The
Perdew-Burke-Ernzerhof exchange-correlation functional
\cite{perdew1996gga} is used in all calculations. This model structure was already used in
\cite{Wang2011} as well as in the first principles TDDFT study
\cite{Yam2011}, so that benchmark data is available for comparison.
Transport through benzenediol is typical for conjugated molecules in
many respects. The transmission at the Fermi energy $E_F$ is rather small
(T $\approx$ 0.01) and transport occurs through the tails of
the $\pi$ and $\pi^*$ frontier orbitals. 

\begin{figure}
  \includegraphics[scale=0.4]{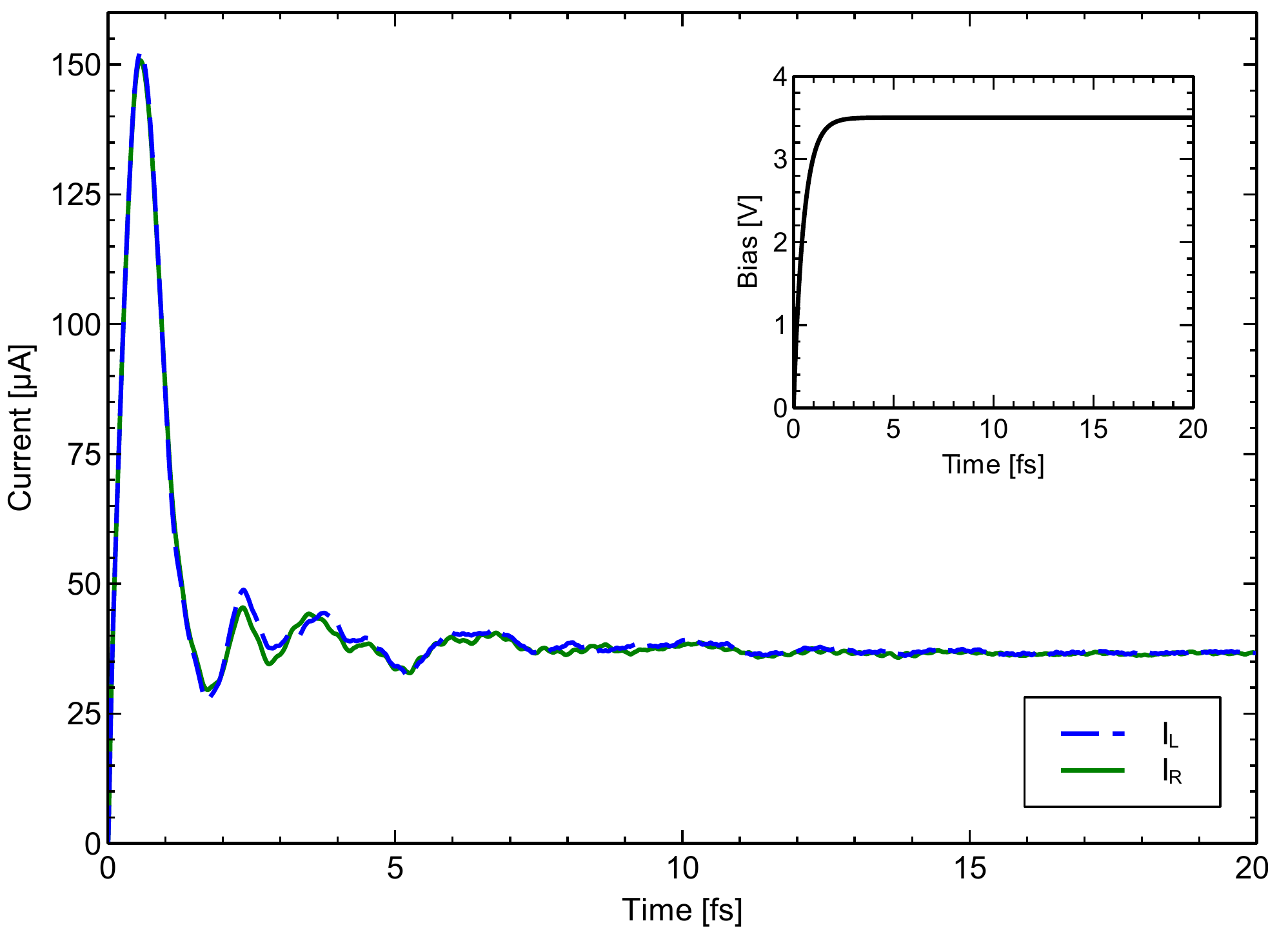}\\
  \caption{Absolute value of the time-dependent current through the left ($I_L$) and right
    ($I_R$) interface of the molecular junction depicted in
    Fig.~\ref{setup}. The inset shows the bias potential $V(t) = V_0
    [1-\exp(-t/T)]$ with $V_0 = 3.5$ V and $T= 0.5$ fs. }\label{Ioft}
\end{figure}

In Fig.~\ref{Ioft} we plot
the time-dependent current through the left and right molecule-lead
interface. Here and in the following we integrate Eq.~\ref{neumann} with a time
step of 2 as using a 4-th order Runge-Kutta method. The bias voltage
of 3.5 V is applied to the left lead only and turned on exponentially with a
time constant of 0.5 fs. The current initially
overshoots, oscillates and settles into the steady state only after
several fs, long after the bias potential nearly reached its
maximum. Earlier we have shown \cite{Wang2011}, that the initial
transients depend on the time constant of the
exponential turn on, but not the asymptotic
value of the current. In addition, we could relate the decay time of
the oscillations to the imaginary part of the self energy of the
device. Well coupled junctions reach the steady state earlier, whereas
weakly coupled junctions feature persistent oscillations (see also
\cite{Khosravi2008}). As can also be seen in Fig.~\ref{Ioft}, the
absolute values of the currents
through the left and right interface equal each other asymptotically,
but not in the transient phase of the simulation. Indeed, the particle
current is a conserved quantity only in the {\em dc} limit. Under {\em
  ac}
driving the device may become charged and one has to consider both
particle current and displacement current \cite{Wang1999}. By
monitoring the total device charge as a function of time, we verified that
the latter indeed compensates for the difference between $|I_L(t)|$ and
$|I_R(t)|$.

\begin{figure}
  \includegraphics[scale=0.4]{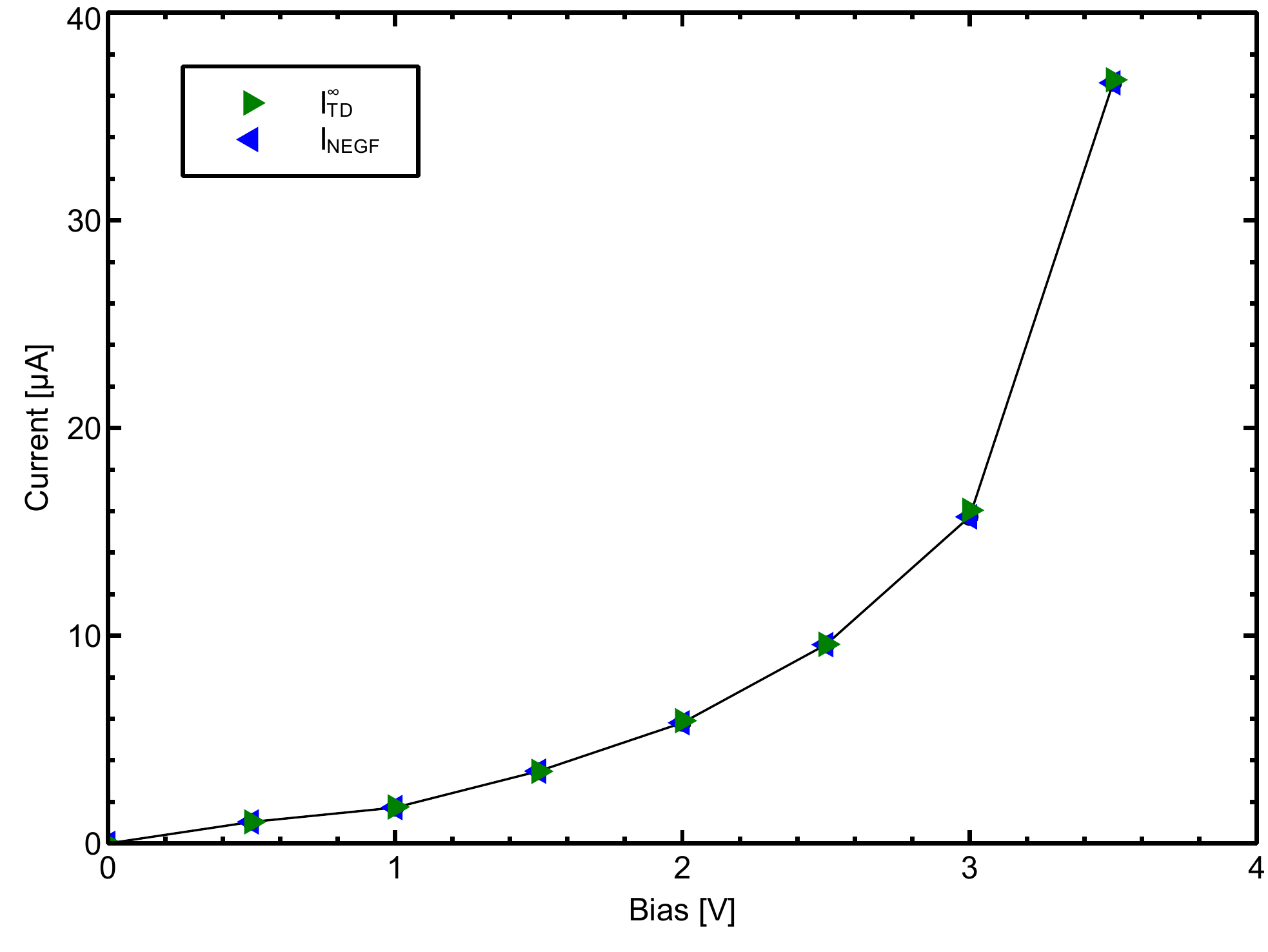}\\
  \caption{Asymptotic time-dependent current ($I_\text{TD}^\infty$)
    and current in the Landauer formalism ($I_\text{NEGF}$) for 1,4-benzenediol as a
    function of applied bias. The values for $I_\text{TD}^\infty$ have
   been obtained from simulations with a total simulation time
   $t_\text{max}$ of
   20 fs and a bias potential $V(t) = V_0
    [1-\exp(-t/T)]$ with  T= 0.5 fs. The current has been averaged
    over the last 2 fs. The wide band approximation was also employed
    in the Landauer calculations. The line is a guide to the eye. }\label{landauer}
\end{figure}

An interesting question is now, whether the asymptotic current
$I_\text{TD}^\infty = \lim_{t \to \infty} I(t)$ from the time-dependent
simulation equals the current obtained from a conventional static
calculation in the Landauer formalism. In the latter approach the
current is given by the energy integral
 \begin{gather}
   I = G_0 \int_{-\infty}^{\infty} dE \left[f(E,\mu_L) -
   f(E,\mu_R)\right]\, T(E,V)\nonumber \\
 T(E,V) =
 {\rm Tr}\left[ G^r
    \Gamma_{R}
    G^a
    \Gamma_{L} \right], \label{t-of-e}
\end{gather} 
with $f(E,\mu)$ denoting Fermi distribution functions with $\mu_L-\mu_R
= V$, the quantum of conductance $G_0
\approx 77.48 \mu$S, and the bias dependent transmission
function $T(E,V)$ \cite{Cuevas2010}. The retarded ($G^r$) and
advanced ($G^a$)
device Green's functions depend on the Hamiltonian and charge
density n({\bf r}). Since n({\bf r}) depends itself on $G^r$ as well
as on the applied bias, a
self-consistent determination of all quantities is required. It is not
{\em a-priori} evident, that the currents given by Eq.~\ref{current}
and Eq.~\ref{t-of-e} are identical. We have recently discussed this
question in great detail in the context of first-principles TDDFT \cite{Yam2011}. Here we perform
similar simulations using the TD-DFTB method in order to show that our
findings are not restricted to a specific choice of the
Hamiltonian. In Fig.~\ref{landauer} we compare the asymptotic
currents $I_\text{TD}^\infty$ from several time-dependent simulations
at different bias values with the corresponding values from
Eq.~\ref{t-of-e}. Despite significant formal and also algorithmic
differences between the two approaches, one can observe nearly
identical values over the full bias range. Like in
Ref.~\cite{Yam2011}, we conclude that time-dependent
  simulations do in general 
  not offer additional or  more accurate information when the interest
  is in steady state properties\footnote{This statement holds for
    conventional local and semi-local functionals of the density. For non-local
    functionals differences with respect to the Landauer approach have
    been predicted \cite{Evers2004,Sai2005,Vignale2009}.}. As we discuss
  in the next section, there is however an important computational
  advantage for {\em ac} transport.   

\subsection{Admittance from current traces}
Starting with the work of Fu  and Dudley \cite{Fu1993}, several
studies addressed the response of meso- and nanoscopic devices to an
alternating bias potential \cite{Jauho1994,Christen1996,Baer2004}. In recent years approaches based on energy
domain Green's functions became especially popular
\cite{Yu2007a,Yamamoto2010,Sasaoka2011,Hirai2011}, but also time
domain techniques, as presented here, allow for the efficient
evaluation of the admittance \cite{Yam2008}. 

To this end, the Fourier transform\footnote{Since $V(t)=0$ for $t<0$, this is
  equivalent to the Laplace transform with imaginary argument.} of bias and current
is numerically evaluated, e.g.,
\begin{equation}
  \label{FT}
  V(\omega) = \int_{-\infty}^\infty V(t) \exp(i\omega t) dt
\end{equation}
to yield the complex admittance $Y(\omega) =
I(\omega)/V(\omega)$. In electronic circuit theory, the real and imaginary parts of $Y$ are also often termed
conductance ($G=\text{Re}(Y)$) and suceptance ($B=\text{Im}(Y)$),
respectively. With the choice for the sign of the Fourier transform
from above (Eq. \ref{FT}), capacitive devices feature a negative
susceptance, while inductive behaviour is characterized by positive
values of $B$.  

\begin{figure}
  \includegraphics[scale=0.4]{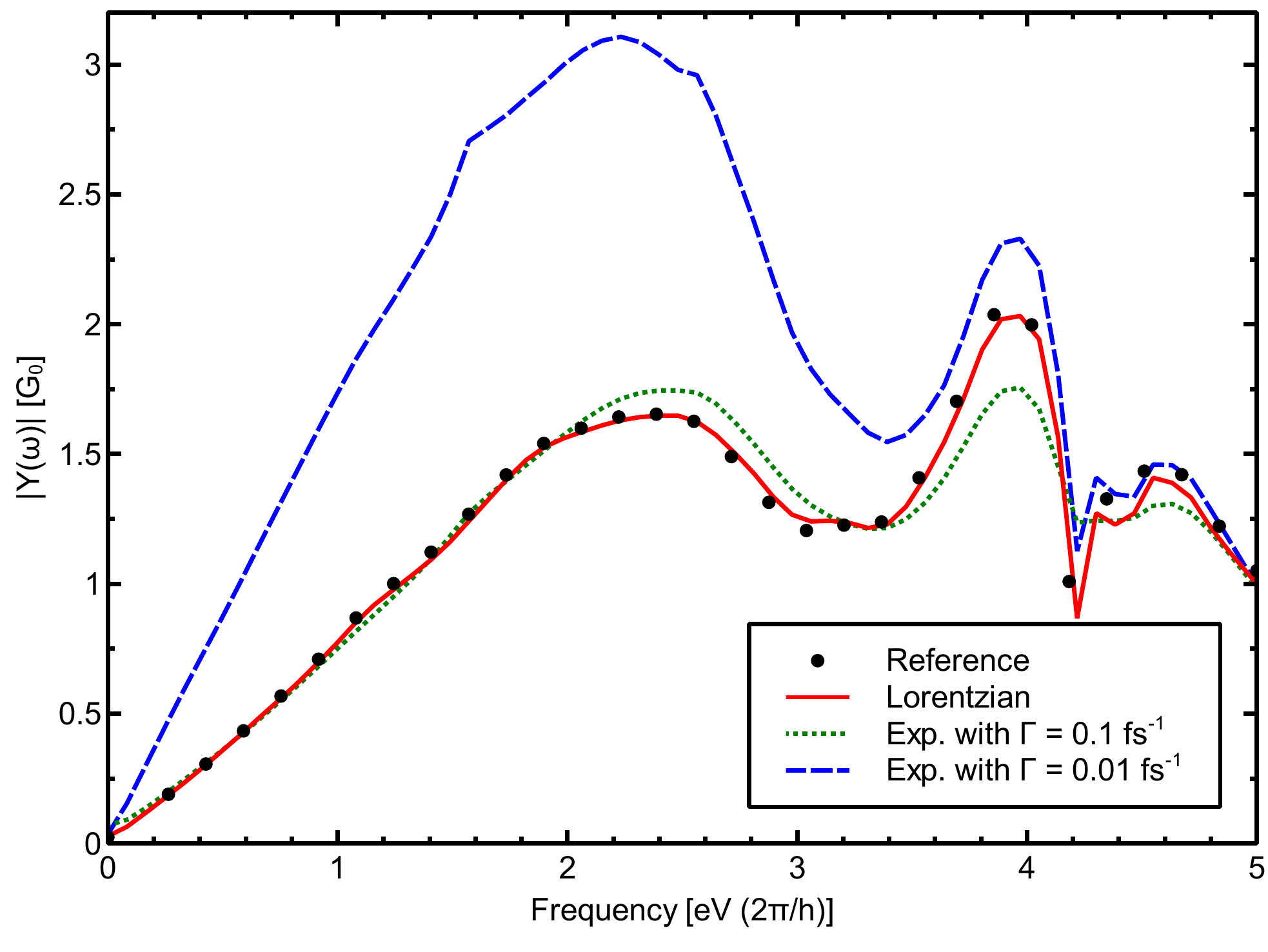}\\
  \caption{Absolute value of admittance $|Y(\omega)|$ in units of
    $G_0$ as a function of frequency in units of [eV/$\hbar$]. Results are given for the  harmonic perturbation
    with discrete frequencies (Reference) and using Fourier
    transforms with exponential form  and damping of the bias ($V_0 =$ 1 mV, $T =$ 0.2 fs,
    $t_\text{max}=$ 50 fs) as well as with Lorentzian form ($V_0 =$
    0.1 mV, $\gamma =$ 0.2 fs, $t_0 =$ 2 fs,
    $t_\text{max}=$ 50 fs).}\label{GofOmega}
\end{figure}

\begin{figure}
  \includegraphics[scale=0.4]{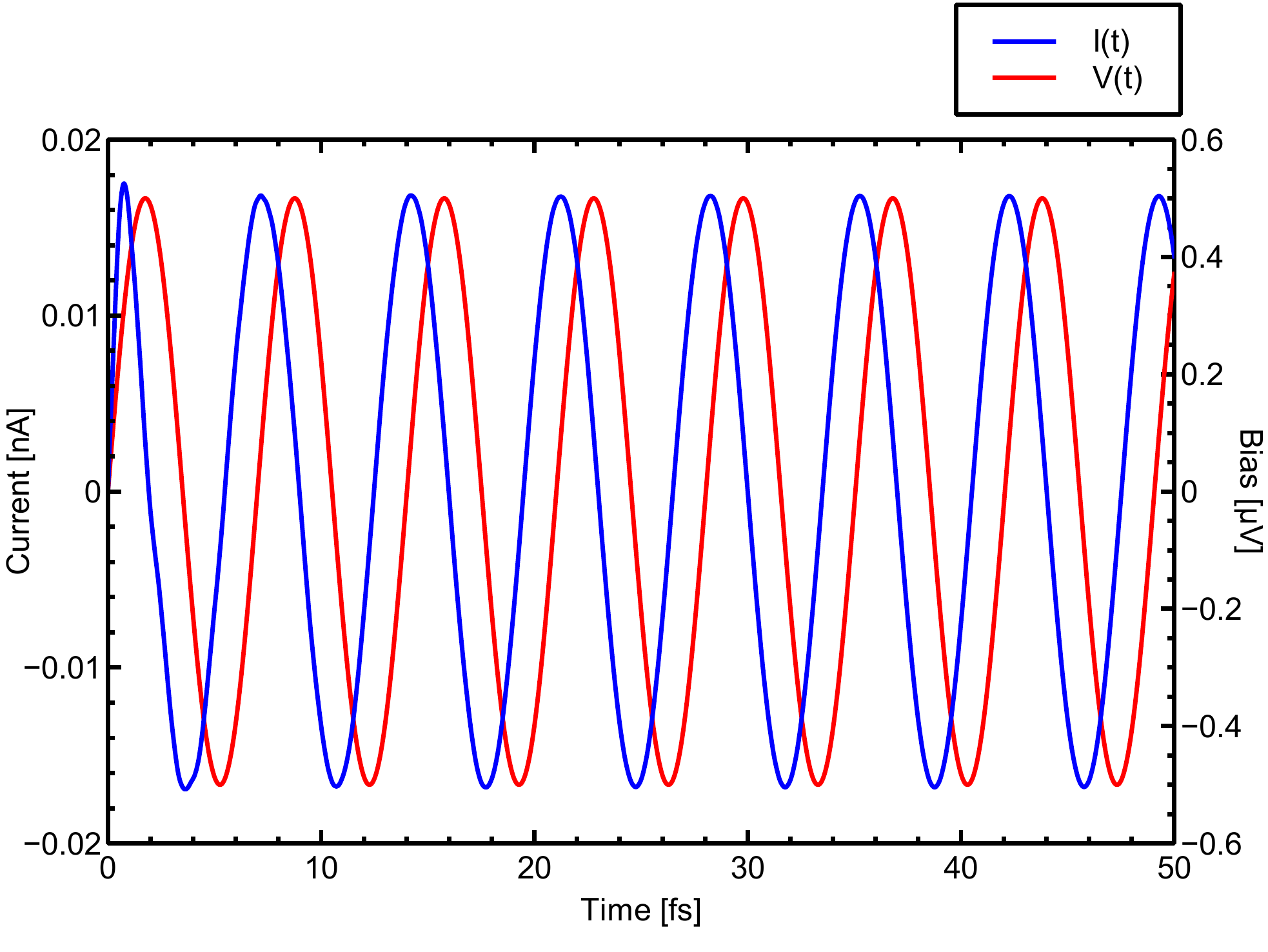}\\
  \caption{time-dependent current due to the {\em ac} bias $V(t) = V_0
    \sin(\omega t)$ ($V_0 =$ 0.5 $\mu$V, $\omega =$ 0.59 eV/$\hbar$,
    $t_\text{max}=$ 50 fs).}\label{sin}
\end{figure}

We applied this approach to the 1,4-benzenediol junction and
experimented with different choices for the temporal profile of the
bias potential. In principle, the form of $V(t)$ is arbitrary as long
as the
amplitude is small enough to remain in the linear response
regime and the support of its Fourier transform is sufficiently large. Fig.~\ref{GofOmega} depicts the absolute value of the
admittance $|Y(\omega)|$ for different functions $V(t)$. As reference, we
perform simulations with a harmonic bias $V(t)= V_0 \sin(\omega t)$
for different discrete values of $\omega$ and determine the amplitude
of $I(t)$ after the initial transients have died out. A sample
simulation is shown in Fig.~\ref{sin}. Inspection of
Fig.~\ref{GofOmega} reveals that an exponential turn-on of the form
$V(t) = V_0 [1-\exp(-t/T)]$ provides a reasonable estimate for the
general features in the admittance, but fails to convince on a
quantitative level. The reason is that the Fourier transform does not
exist in the limit $\omega \to 0$, unless one artificially damps $V(t)$
by a factor $\exp(-\Gamma t)$ to enforce convergence. For small values
of $\Gamma$ the admittance differs strongly from the reference, while
for larger values the {\em dc} limit is overestimated. In response
calculations for optical properties one often uses a Dirac delta
function or Lorentzian as a perturbation.  Here, the Fourier transform exists for
all $\omega$ and no artificial broadening is required. Results for the bias potential 
\begin{equation}
  \label{dirac}
  V(t)= \frac{V_0}{\pi} \frac{\gamma}{(t-t_0)^2 + \gamma^2},
\end{equation}
show excellent agreement with the reference data over nearly the full
frequency range, especially also in the {\em dc} limit. A benefit with
respect to the simulations at discrete frequencies is that the Fourier
transform technique requires only a single run to evaluate
the full admittance. In the following we
therefore continue with this choice. 


\begin{figure}
  \includegraphics[scale=0.35]{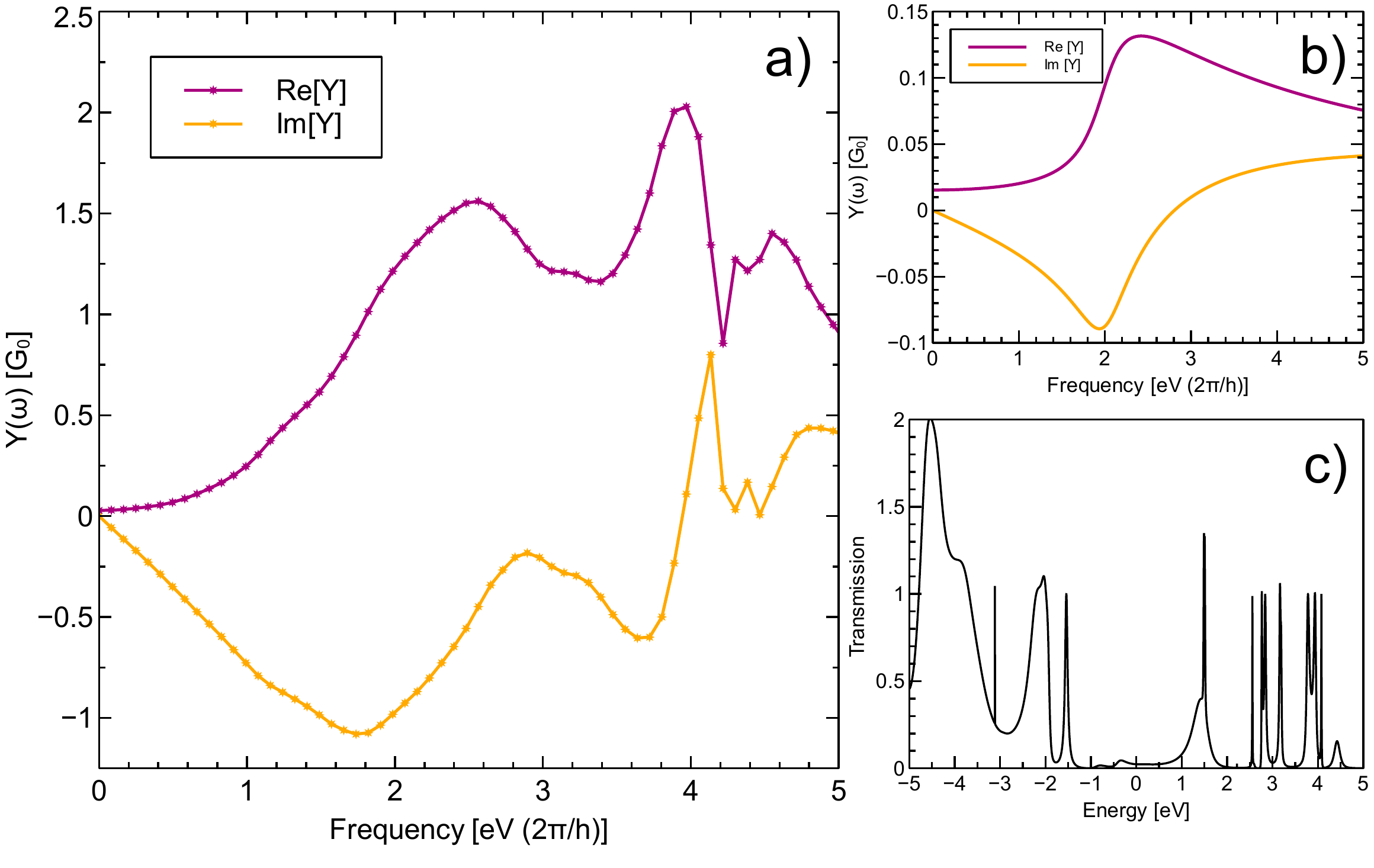}\\
  \caption{a) Real and imaginary part of the admittance
    $Y(\omega)$ for 1,4-benzenediol. b) Analytical results for the one-level model of Fu
    and Dudley \cite{Fu1993} with parameters $\Delta E = $ 2 eV and
    $\gamma=$ 0.25 eV. c) Transmission T(E,0) of 1,4-benzenediol.}\label{AbsIm}
\end{figure}

After this more technical discussion we now analyze the admittance in
more detail. Fig.~\ref{AbsIm} a) shows the conductance and susceptance of
1,4-benzenediol. For small frequencies, the negative values of the
latter indicate capacitive behaviour of the junction. This is in line
with the simulations shown in Fig.~\ref{sin}, where the current leads
the voltage signal. The negative susceptance can be rationalized by
inspection of the transmission T(E,0) (Fig.~\ref{AbsIm} c)) of the
junction. For small frequencies, only the region around the
Fermi energy is relevant in the linear response regime. Here the transmission is low and
the current effectively blocked, similar to a macroscopic
capacitor. In a classical RC circuit, the admittance is given by
\begin{equation}
  Y^\text{RC}(\omega) = -i \omega C + \omega^2 C^2 R,
\end{equation}
up to second order in the frequency \cite{Yu2007a}. As seen in
Fig.~\ref{AbsIm} a), real and imaginary part of $Y(\omega)$ show for
small frequencies indeed quadratic and linear scaling, respectively.     

For larger frequencies, resonances appear around 2.5 eV/$\hbar$ and 4 eV/$\hbar$  with
conductances that are two orders of magnitude larger than the {\em dc}
one. This can be qualitatively understood by a comparison to the
analytical result of Fu and Dudley for a single-level model
\cite{Fu1993}, characterized by a Breit-Wigner transmission 

\begin{equation}
  \label{breit}
  T(E) = \frac{\gamma^2}{(E-E_0)^2 + \gamma^2},
\end{equation}
where $E_0$ denotes the level energy. Here the
admittance is given by

  \begin{multline}
\label{fududr}
      \text{Re}\{Y(\omega)\} = \\G_0 \frac{\gamma}{2\omega} \left[
    \arctan\left(\frac{\Delta E + \omega}{\gamma}\right) -  \arctan\left(\frac{\Delta E - \omega}{\gamma}\right)\right]
  \end{multline}

and 

  \begin{multline}
\label{fududi}
      \text{Im}\{Y(\omega)\} = \\G_0 \frac{\gamma}{4\omega} \ln\left(
        \frac{\left[ (\Delta E + \omega)^2 + \gamma^2 \right]
          \left[ (\Delta E - \omega)^2 + \gamma^2
          \right]}{\left[\Delta E^2  + \gamma^2\right]^2} \right),
  \end{multline}
with $\Delta E = E_F-E_0$. In Fig.~\ref{AbsIm} b) the Fu-Dudley admittance is shown for $\Delta E = $ 2 eV and
$\gamma=$ 0.25 eV. The qualitative similarity to the atomistic results
for a real junction is clearly seen. Our calculations include
a large number of molecular states, while Eqs.~\ref{fududr} and
\ref{fududi} hold only for a single channel. Still a rough assignment of the
resonances to individual molecular states becomes possible.  
The first resonance likely originates from the unoccupied states 1.5 eV
above and the occupied states 2.2 eV below the Fermi energy. According
to the analytical results, states
with smaller broadening contribute less to the admittance. The
resonance around 4 eV is therefore assigned to the broad resonance in the
transmission around this energy. The underlying physical picture is that the
time-dependent bias potential opens new transport channels at E+$\hbar
\omega$ and E -$\hbar \omega$, which are not available in the {\em dc} limit. 

Admittedly, the frequency range for resonant enhancement is difficult to access
experimentally. Current measurements on nanoscopic conductors hardly
reach the GHz regime \cite{Plombon2007,Yamauchi2012}. Nevertheless,
appropriate gating of the device could move the
HOMO/LUMO\footnote{Highest occupied and lowest unoccupied molecular
  orbital.} close to the Fermi energy, resulting in resonance
enhancement at lower frequencies. Small gap materials like Graphene
nanoribbons would offer another route for the experimental
realization of this effect.     

\section{Summary}
In this study, we investigated the behaviour of a contacted
1,4-benzenediol molecule subjected to an alternating bias directly and
using a Fourier transform of Lorentzian and exponential voltage
signals. The Lorentzian input signal led to very good
agreement with reference discrete frequency calculations. In the
admittance, capacitive behaviour could be identified and
interpreted through the transmission of the junction. An analytical
single level model showed large qualitative similarities to our
numerical results. The approach we employed for these findings is a
combination of the highly efficient tight-binding approximation to
adiabatic TDDFT and a device density matrix propagation scheme derived
within the Keldysh formalism by Zheng and co-workers. Following earlier
discussions on this topic, we can also confirm that a different choice
of the self-consistent Hamiltonian does not change the equality of the TD steady state
with its static counterpart at the NEGF level.

\begin{acknowledgement}
Financial support by the German Science Foundation (DFG, SPP 1243 and
GRK 1570) is greatly acknowledged.
\end{acknowledgement}

\bibliographystyle{elsarticle-num}
\bibliography{../../Combined}

\begin{thebibliography}{10}
\expandafter\ifx\csname url\endcsname\relax
  \def\url#1{\texttt{#1}}\fi
\expandafter\ifx\csname urlprefix\endcsname\relax\def\urlprefix{URL }\fi
\expandafter\ifx\csname href\endcsname\relax
  \def\href#1#2{#2} \def\path#1{#1}\fi

\bibitem{Nitzan2003}
A.~Nitzan, M.~A. Ratner, Electron transport in molecular wire junctions,
  Science 300~(5624) (2003) 1384--1389.

\bibitem{Heath2009}
J.~R. Heath, Molecular electronics, Annu. Rev. Mater. Res. 39 (2009) 1--23.

\bibitem{McCreery2013}
R.~L. McCreery, H.~Yan, A.~J. Bergren, A critical perspective on molecular
  electronic junctions: there is plenty of room in the middle, Phys. Chem.
  Chem. Phys. 15~(4) (2013) 1065--1081.

\bibitem{Bogani2008}
L.~Bogani, W.~Wernsdorfer, Molecular spintronics using single-molecule magnets,
  Nat. Mater. 7~(3) (2008) 179--186.

\bibitem{Song2009}
H.~Song, Y.~Kim, Y.~H. Jang, H.~Jeong, M.~A. Reed, T.~Lee, Observation of
  molecular orbital gating, Nature 462~(7276) (2009) 1039--1043.

\bibitem{Reddy2007}
P.~Reddy, S.-Y. Jang, R.~A. Segalman, A.~Majumdar, Thermoelectricity in
  molecular junctions, Science 315~(5818) (2007) 1568--1571.

\bibitem{Nikolic2012}
B.~K. Nikoli{\'c}, K.~K. Saha, T.~Markussen, K.~S. Thygesen, First-principles
  quantum transport modeling of thermoelectricity in single-molecule
  nanojunctions with graphene nanoribbon electrodes, Journal of Computational
  Electronics 11 (2012) 1--15.

\bibitem{Gagliardi2008}
A.~Gagliardi, G.~Romano, A.~Pecchia, A.~Di~Carlo, T.~Frauenheim, T.~A. Niehaus,
  Electron-phonon scattering in molecular electronics: from inelastic electron
  tunnelling spectroscopy to heating effects, New J. Phys. 10 (2008) 065020.

\bibitem{Schulze2008}
G.~Schulze, K.~J. Franke, A.~Gagliardi, G.~Romano, C.~Lin, A.~Da~Rosa, T.~A.
  Niehaus, T.~Frauenheim, A.~Di~Carlo, A.~Pecchia, J.~I. Pascual, Resonant
  electron heating and molecular phonon cooling in single c$_{60}$ junctions,
  Phys. Rev. Lett. 100 (2008) 136801.

\bibitem{Cuevas2010}
J.~C. Cuevas, E.~Scheer, Molecular Electronics: An Introduction to Theory and
  Experiment, World Scientific, 2010.

\bibitem{Chen2012}
G.~Chen, T.~Niehaus, Quantum Simulation for Material and Biological systems,
  Springer, 2012, Ch. Quantum transport simulations based on time dependent
  density functional theory, pp. 17--32.

\bibitem{Niehaus2009}
T.~A. Niehaus, Approximate time-dependent density functional theory, J. Mol.
  Struct.: THEOCHEM 914 (2009) 38.

\bibitem{Wang2011}
Y.~Wang, C.-Y. Yam, T.~Frauenheim, G.~Chen, T.~Niehaus, An efficient method for
  quantum transport simulations in the time domain, Chem. Phys. 391~(1) (2011)
  69.

\bibitem{Yam2011}
C.~Y. Yam, X.~Zheng, G.~H. Chen, Y.~Wang, T.~Frauenheim, T.~A. Niehaus,
  Time-dependent versus static quantum transport simulations beyond linear
  response, Phys. Rev. B 83 (2011) 245448.

\bibitem{Zheng2007}
X.~Zheng, F.~Wang, C.~Y. Yam, Y.~Mo, G.~H. Chen, Time-dependent
  density-functional theory for open systems, Phys. Rev. B 75~(19) (2007)
  195127.

\bibitem{Casida1995}
M.~E. Casida, Recent Advances in Density Functional Methods, Part I, World
  Scientific, 1995, Ch. Time-dependent Density Functional Response Theory for
  Molecules, pp. 155--192.

\bibitem{Frauenheim2002}
T.~Frauenheim, G.~Seifert, M.~Elstner, T.~Niehaus, C.~Kohler, M.~Amkreutz,
  M.~Sternberg, Z.~Hajnal, A.~Di~Carlo, S.~Suhai, Atomistic simulations of
  complex materials: ground-state and excited-state properties, Journal Of
  Physics-Condensed Matter 14~(11) (2002) 3015--3047.

\bibitem{perdew1996gga}
J.~Perdew, K.~Burke, M.~Ernzerhof, Generalized gradient approximation made
  simple, Phys. Rev. Lett. 77~(18) (1996) 3865--3868.

\bibitem{Khosravi2008}
E.~Khosravi, S.~Kurth, G.~Stefanucci, E.~K.~U. Gross, The role of bound states
  in time-dependent quantum transport, Appl. Phys. A 93~(2) (2008) 355--364.

\bibitem{Wang1999}
B.~Wang, J.~Wang, H.~Guo, Current partition: A nonequilibrium green's function
  approach, Phys. Rev. Lett. 82~(2) (1999) 398--401.

\bibitem{Evers2004}
F.~Evers, F.~Weigend, M.~Koentopp, Conductance of molecular wires and transport
  calculations based on density-functional theory, Phys. Rev. B 69~(23) (2004)
  235411.

\bibitem{Sai2005}
N.~Sai, M.~Zwolak, G.~Vignale, M.~Di~Ventra, Dynamical corrections to the
  dft-lda electron conductance in nanoscale systems, Phys. Rev. Lett. 94~(18)
  (2005) 186810.

\bibitem{Vignale2009}
G.~Vignale, M.~Di~Ventra, Incompleteness of the landauer formula for electronic
  transport, Phys. Rev. B 79~(1) (2009) 14201.

\bibitem{Fu1993}
Y.~Fu, S.~Dudley, Quantum inductance within linear response theory, Phys. Rev.
  Lett. 70~(1) (1993) 65--68.

\bibitem{Jauho1994}
A.~Jauho, N.~Wingreen, Y.~Meir, Time-dependent transport in interacting and
  noninteracting resonant-tunneling systems, Phys. Rev. B 50~(8) (1994) 5528.

\bibitem{Christen1996}
T.~Christen, M.~B\"uttiker, Low frequency admittance of a quantum point
  contact, Phys. Rev. Lett. 77 (1996) 143--146.

\bibitem{Baer2004}
R.~Baer, T.~Seideman, S.~Ilani, D.~Neuhauser, Ab initio study of the
  alternating current impedance of a molecular junction, J. Chem. Phys. 120
  (2004) 3387.

\bibitem{Yu2007a}
Y.~Yu, B.~Wang, Y.~Wei, Corrected article: ``ac response of a carbon chain
  under a finite frequency bias'' [j. chem. phys. [bold 127], 104701 (2007)],
  J. Chem. Phys. 127~(16) (2007) 169901.

\bibitem{Yamamoto2010}
T.~Yamamoto, K.~Sasaoka, S.~Watanabe, Universal transition between inductive
  and capacitive admittance of metallic single-walled carbon nanotubes, Phys.
  Rev. B 82~(20) (2010) 205404.

\bibitem{Sasaoka2011}
T.~Sasaoka, K.and~Yamamoto, S.~Watanabe, K.~Shiraishi, ac response of quantum
  point contacts with a split-gate configuration, Phys. Rev. B 84 (2011)
  125403.

\bibitem{Hirai2011}
D.~Hirai, T.~Yamamoto, S.~Watanabe, Theoretical analysis of ac transport in
  carbon nanotubes with a single atomic vacancy: Sharp contrast between dc and
  ac responses in vacancy position dependence, Applied Physics Express 4~(7)
  (2011) 075103.

\bibitem{Yam2008}
C.~Y. Yam, Y.~Mo, F.~Wang, X.~Li, G.~H. Chen, X.~Zheng, Y.~Matsuda,
  J.~Tahir-Kheli, A.~G. William~III, Dynamic admittance of carbon
  nanotube-based molecular electronic devices and their equivalent electric
  circuit, Nanotech. 19 (2008) 495203.

\bibitem{Plombon2007}
J.~Plombon, K.~P. OBrien, F.~Gstrein, V.~M. Dubin, Y.~Jiao, High-frequency
  electrical properties of individual and bundled carbon nanotubes, Appl. Phys.
  Lett. 90~(6) (2007) 063106--063106.

\bibitem{Yamauchi2012}
K.~Yamauchi, S.~Kurokawa, A.~Sakai, Admittance of au/1, 4-benzenedithiol/au
  single-molecule junctions, Appl. Phys. Lett. 101~(25) (2012) 253510--253510.

\end{thebibliography}
\end{document}